\documentclass[twocolappendix]{emulateapj}
\slugcomment{}

\usepackage{amsmath}
\usepackage{natbib}
\usepackage[para]{threeparttable}

\begin{document}
\shortauthors{Shah \& Nelemans}
\shorttitle{Tidal astrophysics of white-dwarfs}

\title{Measuring tides and binary parameters from gravitational wave
  data and eclipsing timings of detached white dwarf binaries}

\author{Sweta Shah$^{1,2}$ and Gijs Nelemans$^{1,2,3}$}

\affil{ $^1$Department of Astrophysics/ IMAPP, Radboud University Nijmegen,
  P.O. Box 9010, 6500 GL Nijmegen, The Netherlands\\
	$^2$Nikhef – National Institute for Subatomic Physics,  Science
  Park 105,  1098 XG Amsterdam, The Netherlands \\
	$^3$Institute for Astronomy, KU Leuven, Celestijnenlaan 200D, 3001
  Leuven, Belgium\\	}

\email{s.shah@astro.ru.nl}
\date{\today}
\begin{abstract} 
The discovery of the most compact detached white dwarf (WD) binary
  SDSS J065133.33+284423.3 has been discussed in terms of probing the 
  tidal effects in white dwarfs. This system is also a \emph{verification
  source} for the space-based gravitational wave (GW) detector,
evolved \emph{Laser Interferometer Space Antenna} (eLISA) which will
observe short-period compact Galactic binaries with $P_{\mathrm{orb}}
\lesssim 5$ hrs. We address the prospects of doing tidal studies using
eLISA binaries by showing the fractional uncertainties in the orbital
decay rate and the rate of that decay, $\dot{f}, \ddot{f}$ expected
from both the GW and EM data for some of the high-$f$ binaries. We
find that $\dot{f}$ and $\ddot{f}$ can be measured using GW data only
for the most massive WD binaries observed at high-frequencies. Form
timing the eclipses for $\sim 10$ years, we find that $\dot{f}$ can be
known to $\sim 0.1\% $ for J0651. We find that from GW data alone,
measuring the effects of tides in binaries is (almost) impossible. We
also investigate the improvement 
in the knowledge of the binary parameters by combining GW amplitude
and inclination with EM data with and without $\dot{f}$. In our
previous work we found that EM data on distance constrained 2-$\sigma$
uncertainty in chirp mass to $15-25\%$ whereas adding $\dot{f}$
reduces it to $0.11\%$. EM data on $\dot{f}$ also constrains
2-$\sigma$ uncertainty in distance to $35\%-19\%$. EM data on primary
mass constrains the secondary mass $m_2$ to factors of 2 to $\sim
40\%$ whereas adding $\dot{f}$ reduces this to $25\%$. And
finally using \emph{single-line spectroscopic} constrains 2-$\sigma$
uncertainties in both the $m_2, d$ to factors of 2 to $\sim
40\%$. Adding EM data on $\dot{f}$ 
reduces these 2-$\sigma$ uncertainties to $\leq 25\%$ and  $6\%-19\%$ 
respectively. Thus we find that EM measurements of $\dot{f}$ and
radial velocity will be valuable in constraining eLISA binary parameters.
\end{abstract}

\keywords{stars: white-dwarfs - tides, binaries - gravitational waves,
  Galactic binaries - parameters, GW detectors - eLISA}
\maketitle
\section{Introduction}
The discovery of a detached white
dwarf (WD) eclipsing binary system, SDSS J065133.33+284423.3 (J0651,
hereafter) \citep{2011ApJ...737L..23B} has generated a number of
discussions on the subject of tidal physics of compact objects
\citep[e.g.][]{2011ApJ...740L..53P, 2012MNRAS.421..426F,
  2013MNRAS.tmp..614F, 2013MNRAS.tmp.1449B, 2013MNRAS.428..518D}. The
small orbital period of $P_{\mathrm{orb}} \approx 765$s, and the compact
nature of the stars which are not yet transferring mass, makes it one of
the most interesting candidates for studying the level of tidal
interactions between the components and the possible astrophysical
implications for WDs. J0651 is also a \textit{verification binary}
for eLISA\footnote{a space-based gravitational wave mission with
  expected launch in 2034} \citep{2013GWN.....6....4A} radiating
gravitational wave at $f = 2.6$mHz  with an estimated signal-to-noise
ratio (S/N) of $\sim$10 for an observation length of 2 years. In this
paper, we investigate detectability of the tidal effects from the GW
and EM data and their implications on the astrophysical knowledge of
the WDs in the binary and similar systems. In addition we discuss the
effect of using measured EM period changes on the GW parameter
estimates. 

eLISA will observe Galactic binaries with periods shorter than a
few hours. While the majority of the binaries (which are mostly double WD
objects) are radiating GWs in the low-frequency range ($f \leq 3$mHz),
there are a handful of high-frequency sources with significant orbital
decay as predicted by population synthesis simulations
\citep{2004MNRAS.349..181N}. Despite
the limited number of such high-$f$ objects, they 
present a unique opportunity to do tidal studies of compact objects
as these  relatively high-$f$ binaries will have a strong gravitational signal
strength and larger values for their rate of change of the orbital
periods both aiding accurate GW measurements of their orbital
parameters. Here, we use Fisher studies\citep{1998PhRvD..57.7089C} to address the 
detectability of the rate of change of the source's GW frequency,
$\dot{f}$ and $\ddot{f}$ from the GW data for the detached J0651-like
binary systems. The GW parameters, $f$,
$\dot{f}$ and $\ddot{f}$ of a circular binary are trivially related to the more familiar
quantities in EM observations, $P_{\mathrm{orb}}$,
$\dot{P}_{\mathrm{orb}}$, and $\ddot{P}_{\mathrm{orb}}$ via: $f =
2/P_{\mathrm{orb}}$, $\dot{f}= -2 \dot{P}_{\mathrm{orb}}/P_{\mathrm{orb}}^2$, $\ddot{f} =
2(\dot{P}_{\mathrm{orb}}^2\: -
\:2\:P_{\mathrm{orb}}\: \ddot{P}_{\mathrm{orb}})/P_{\mathrm{orb}}^3$. 

As a compact binary ages via GW dissipation, the orbital period
changes as a result of increasing $\dot{f}$. If the stellar components
in the binary are close enough to each other, an additional source of
dissipation of orbital energy can ensue through tides and this may
reflect in its GW \emph{phase shift}. In this paper we
consider only detached WD systems where both the GW emission and tidal
torque (including dynamical tides) can enhance the orbital decay
rate. The orbital evolution in the presence of mass transfer and GW
(see Eq.~12 in \cite{2004MNRAS.349..181N}) competes with dissipations
from the tides. In these cases, their orbital evolution can be influenced by
short-term variations like the nova explosions and this could
dramatically increase $\dot{f}, \ddot{f}$. This means that in the
millions of binaries that eLISA will observe, if a number of such 
(mass-transferring) systems undergo such orbital perturbation, their
$\dot{f}, \ddot{f}$ will increase by orders of magnitude making it
possible to measure them, however this is very unlikely in the
lifetime of eLISA \citep{2009MNRAS.400L..24S}.
 
Recent studies using EM data have shown that for  the case of J0651,
the period change can be enhanced by roughly up to $5\%$ due to the
tides\citep{2013MNRAS.tmp.1449B, 2011ApJ...740L..54B,
  2011ApJ...740L..53P}. Based on parametrized equilibrium tide theory,
\citet{2011ApJ...740L..53P} has shown that for the J0651 system, in
addition to the GW radiation, the tidal interactions between the WDs
will imprint a shift in the time of eclipses by 0.3s after one year of
timing. The dominant GW contribution advances the shift by
5.5s. \citet{2011ApJ...740L..54B} also calculated the
deviation from the pure GR-driven inspiral, under the assumption that
the WDs are tidally locked with the orbit and the GW radiation causes
a small mismatch between the WD spin and orbital period. This causes a
tidal distortion of the lower mass WD and assuming that this tidal
energy is mostly transferred from orbit to the spin keeping the system 
tidally locked, the tidal deviations were computed for J0651. Both of these
works are corroborated 
by\citet{2013MNRAS.tmp.1449B} who compute the tidal response of
J0651-like system assuming that both WDs are in resonance lock where
the orbit and spin vary uniformly. It has been further claimed that
for J0651, one should be able to detect the effect of  tides in the GW
phase shifts \citep{2012MNRAS.421..426F}. These results are based on
modeling dynamical tide in a carbon/oxygen WD. The prospect of
detecting such a phase shift in the GW data is very exciting as this
could lead to measurements of the components' moment of
inertia. However in order for the tides to significantly shift in the
collective phase of the GW signal, one needs to observe the system for
millions of cycles according to the estimate of the evolution in the
number of cycles only due to the tides
\citep[Eq. 88,][]{2012MNRAS.421..426F} which is not feasible with
currently planned eLISA mission. 
\begin{table*}[!t]
\centering
\begin{threeparttable}
\caption{GW parameter values of J0651, high-mass and the high-$f$
  binary systems}
\label{tab:verf_bin}
\begin{tabular}{c c c c c c c c c c c c }
\hline \hline
&$\mathcal{A}[\times 10^{-22}]$ & $\phi_0$[rad] & $\cos\iota$ & $f[\times10^{-3}]$[Hz] & $\dot{f}$[Hz/s]& $\ddot{f}$[Hz/s$^2$]&$\psi$[rad] & $\sin\beta$ & $\lambda$[rad] & $\mathrm{S/N} $ \\
\hline
J0651 &$1.67$\tnote{a} & $\pi$ & $0.007$ & $2.61$ &
$-3.35\times10^{-17}$ & $1.57\times10^{-31}$ & $\pi/2$ & $0.101$
&$1.77 $ & $\sim 13$\tnote{a} \\
high-mass &$6.71$\tnote{b} & $\pi$ & $0.007$ & $2.61$ &
$-1.07\times10^{-16}$ & $1.61\times10^{-29}$ & $\pi/2$ & $0.101$
&$1.77 $ &$\sim 50$\tnote{b} \\
high-$f$\tnote{c} &$3.69$ & $5.41$ & $-0.86$ & $17.69$ & $-1.99\times10^{-13}$ & $8.19\times10^{-23}$ & $0.75$ & $0.94$ & $1.97$ & $\sim 135$ \\
\hline
      \end{tabular}
    \begin{tablenotes}
        \item[a]for $m_1 = 0.25 M_{\odot}$, $m_2 = 0.55 M_{\odot}, d = 1.0$ kpc
        \item[b] high-mass system with $m_1, m_2 = 0.8 M_{\odot}$, $d = 1.0$ kpc 
        \item[c]  For the given $f, \dot{f}, \ddot{f}$, $m_1, m_2 = 1.01 M_{\odot}$, $d = 9.95$ kpc 
\end{tablenotes}
\end{threeparttable}
\end{table*}

In order to investigate the measurability of the above-mentioned
orbital parameters, we calculate the predicted GW uncertainties in
those parameters as a function of orbital period. We summarize the
data analysis and the selection of the binaries in Section 2. In
Section 3 we estimate the expected EM uncertainties from mid-eclipse
timing measurements. This is followed by a comparison of the
accuracies from two types of measurements in Section 4. Finally we
summarize prospects of measuring deviation in evolution due to tides
and the improvement in the knowledge of the WDs from combining
the accuracies of GW and EM measurements from the 
measurement of the rate of change of orbital period.
\section{ \lowercase{e}LISA binaries and uncertainties from the GW data}
We obtain the GW accuracies by carrying out Fisher information
matrix (FIM) calculations in order to determine whether the GW
parameters $\dot{f}, \ddot{f}$ can be measured over the two year GW
observations by eLISA mission. We consider three binary systems for
this purpose: the verification source J0651, a hypothetical high-mass
J0651 system and the highest-$\dot{f}$ source we find in the 
population synthesis predictions \citep{2004MNRAS.349..181N}. In
the rest of the paper we will 
refer to them as J0651, high-mass and high-$f$ systems
respectively. We list the GW parameter values 
of all these systems in Table I. For J0651 only ${P}_{\mathrm{orb}}$,
and $\dot{P}_{\mathrm{orb}}$ are measured
\citep{2012ApJ...757L..21H}. These have been converted to $f$, and
$\ddot{f}$ with relations mentioned above. Since 
$\ddot{P}_{\mathrm{orb}}$ is not yet measured a fiducial
$\ddot{f}$ has been chosen such that it agrees with the GR
predictions. These values are slightly higher for the high-mass J0651
in accordance with the masses. For the high-$f$ system, the values $\dot{f},
\ddot{f}$ are given by the simulation. 

Our method and application of  FIM to extract the GW parameter uncertainties has
been described in detail in \cite{2012A&A...544A.153S}. In this paper, we extend our 
previous FIM analyses to include \emph{nine} GW parameters:
dimensionless amplitude ($\mathcal{A}$), frequency ($f$), polarisation
angle ($\psi$), initial GW phase ($\phi_{0}$), inclination ($\cos
\iota$), ecliptic latitude ($\sin \beta$), ecliptic longitude
($\lambda$), orbital decay rate ($\dot{f}$), and rate of change of
that decay ($\ddot{f}$). Given these (GW) parameters, we calculate a
$9\times 9$ FIM for all three systems. This 
implies not knowing any of the parameters \textit{a priori}. By
inverting this matrix we get the variance covariance matrix (VCM)
which provides the uncertainties in the parameters and the
correlations between them. We refer to our previous paper for 
the signal and noise modeling in computing the expected 
parameter uncertainties and the correlations between them.
We list the full VCM matrices for J0651, and the high-$f$
systems in the Appendix, which include the normalized correlations
between the 9 parameters. The normalized correlations between
parameters of J0651 and high-$f$ are different because of the
difference between their angular parameters (see
\cite{2013A&A...553A..82S}) and also due to their 
respective GW frequencies \citep{2011PhRvD..83h3006B}. 
\section{Uncertainties from the EM data}
In this section we describe the prospects of extracting the
uncertainties in $f$, $\dot{f}$ and $\ddot{f}$ from the
electromagnetic data. J0651 has a measured $\mathrm{\dot{P}_{orb}} =
−9.8 \pm 2.8 \times 10^{-12} $ s s$^{-1}$ which is consistent with GR
predictions \citep{2012ApJ...757L..21H} within the error. The
way this is typically measured is to compare the observed (O)
mid-eclipse times with computed (C) values from a model with constant
orbital period and fit the O-C values as function of time
\citep[e.g.][]{1991ApJ...378L..45K}. A possible resulting parabola
gives an evidence of a finite value of $\mathrm{\dot{P}_{orb}}$
\citep{2005ASPC..335....3S}. The phase of the signal in cycles at an
arbitrary time $t$ after a reference time evolves and it is given by a
Taylor expansion of the phase: 
\begin{equation}
\phi = \phi_0 \:+ \: f \: (t-t_0) +\frac{\dot{f}}{2} \:(t-t_0)^2\:+\:\frac{\ddot{f}}{6}\: (t-t_0)^3 \:+ \: ... \: , 
\label{eq:phase}
\end{equation}
where $t_0$ is the epoch, and $t$ is measured in the barycentric
co-ordinates. As the source is observed for a longer time, the second and
third terms gain significance. Given a duration of observation,
$\mathrm{T_{obs}}$ and for a fixed resolution in phase ($\sigma_{\phi}$), the
uncertainties in the three orbital parameters can be estimated by 
\citep{1998ApJ...493..891M}:
\begin{equation}
\sigma_f \sim \frac{\sigma_{\phi}}{\mathrm{T_{obs}}}  \:\:\:
;\:\:\:\sigma_{\dot{f}} \sim 2 \frac{\sigma_{\phi}}{\mathrm{T_{obs}}^2}
\:\:\: ;   \:\:\:\sigma_{\ddot{f}} \sim
6 \frac{\sigma_{\phi}}{\mathrm{T_{obs}}^3}       
\label{eq:predicted_error}
\end{equation}
Considering an uncertainty of eclipse timing for J0651 of 
\citet{2012ApJ...757L..21H} $\sigma_{T_0} \sim 0.725$s (see Table 2) gives a
fractional phase error of $\sigma_{T_0}/P_0 \sim 9.5 \times 10^{-4}$
turns. Assuming a constant phase error timing this source for a long
time, for e.g. $\mathrm{T_{obs}} \sim 10$ years using the
above equation we get, $\sigma_f \sim 10^{-12}$Hz, $\sigma_{\dot{f}}
\sim 10^{-21}$Hz/s and $\sigma_{\ddot{f}} \sim 10^{-30}$Hz/s$^2$. This
implies for J0651 the relative uncertainties are $\sigma_{\dot{f}}/\dot{f}
\sim 10^{-5}$, $\sigma_{\ddot{f}}/\ddot{f} \sim 6$. Thus timing J0651
will be very useful to pin down the rate of 
change of frequency, however the uncertainty in $\ddot{f}$ is very large.
Below we will compare the uncertainties in decay rate and rate of the
decay for all three binaries using GW and EM observations for a range
of orbital periods. 
\section{Measurability of \lowercase{$\dot{f}$, $\ddot{f}$}}
\begin{figure*}
\centering 
\includegraphics[width=18cm]{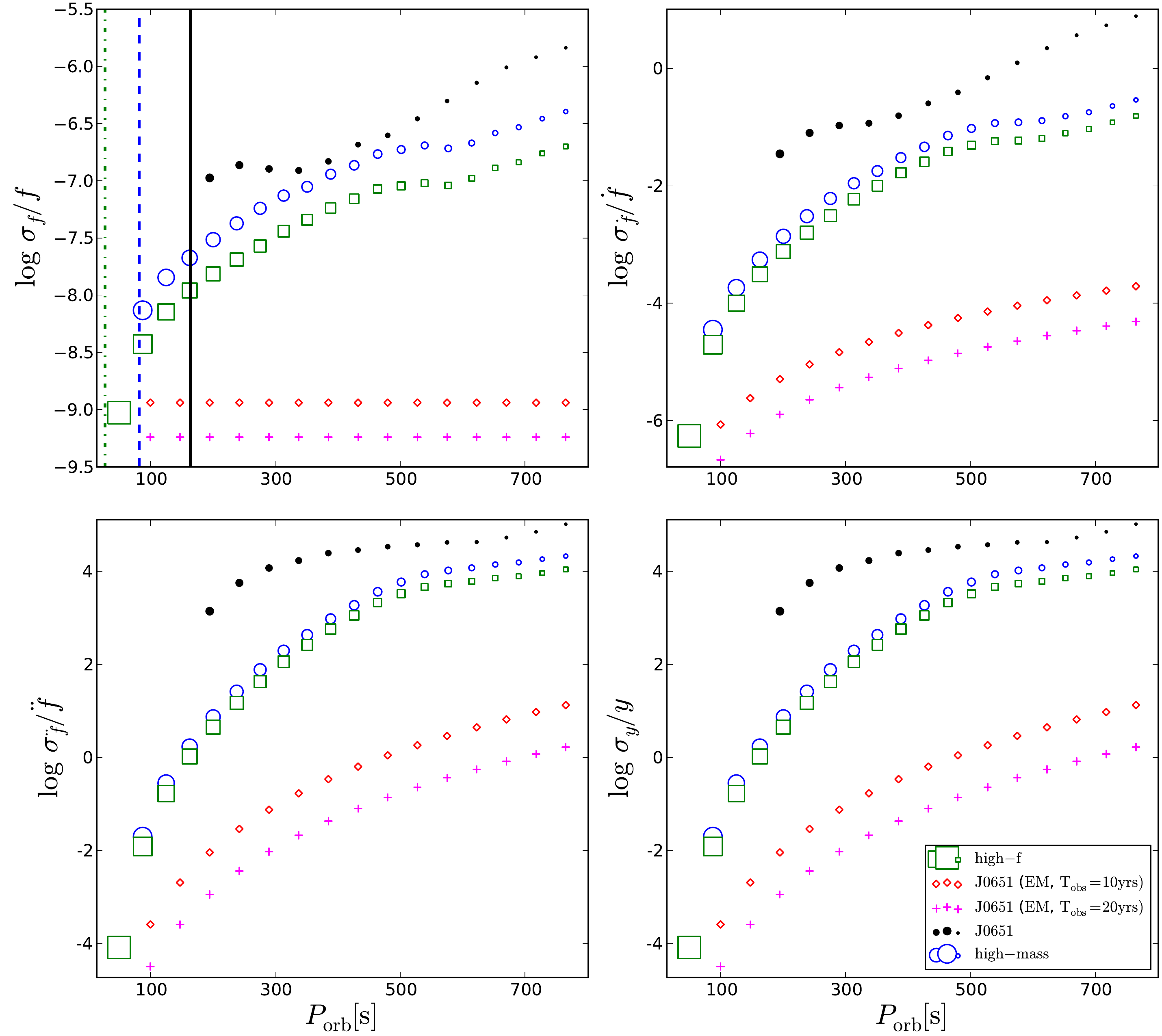}
\caption{Relative uncertainties in frequency ($f$), decay rate ($\dot{f}$), rate of
  decay rate ($\ddot{f}$) and braking index ($y$) using gravitational
  wave data of J0651, the high-mass counterpart and the high-$f$
  binary. All the GW uncertainties are represented by (black) filled
  circle, (blue) open circle and (green) square for the three binary
  systems respectively listed above. Also the same uncertainties are
  shown for J0651 using its electromagnetic observations of its
  eclipse timings which are represented by (red) diamonds for
  $\mathrm{T_{obs}} = 10 \mathrm{yrs}$ and by (magenta) crosses for
  $\mathrm{T_{obs}} = 20 \mathrm{yrs}$. The marker sizes of 
  filled/open circles and open squares represent the signal-to-noise
  ratio from the GW data of each system at that orbital period. The
  vertical lines in the top-left panel are the values of minimum
  orbital period at which a given system will start mass transfer. The
(black) solid line is for J0651, (blue) dashed line is for the
high-mass counterpart and the (green) dash-dotted line is for the
high-$f$ binary.}
\label{fig:GW_errors}
\end{figure*}
A straight forward way to distinguish the tidal contribution from that of 
the GW radiation in the evolution of the binary is to measure the
quantities $f, \dot{f}, \ddot{f}$ with sufficient accuracy. The
general relativistic predictions of the orbital decay in a binary
orbit due to GW radiation \emph{alone} gives the following relation
\citep{1998AIPC..456...61W}: 
\begin{equation}
\left(\frac{\ddot{f}\:f} {{\dot{f}}^2}\right)_{\mathrm{GW}}  \: := y \:= \:\frac{11}{3},
\label{eq:tides}
\end{equation}
thus, a measure of any deviation from this numerical value measured
within the parameter accuracies for detached binaries will provide a
testbed for effects of the tides. 

To get a rough estimate of the percentage of tidal contribution in the
binary evolution of J0651-like systems, we can estimate
Eq.\,~\ref{eq:tides} for J0651 where the tidal contribution is taken
into account since the individual masses and radii of this system have
been measured from its light curve. This gives us an idea of what the
uncertainties in $\dot{f}, \ddot{f}$ must be in order to measure any
deviation from the GR driven binary evolution. Under the influence of GW
radiation only, the rate of change of GW frequency changes according to 
\begin{equation}
\dot{f}_{0} = \frac{96 \:
  \pi}{5}\frac{G^{5/3}}{c^5}\:(\pi\;\mathcal{M}_c)^{5/3}\:{f}^{11/3}, 
\label{eq:fdot}
\end{equation}
where $\mathcal{M}_c $ is the chirp mass given by:
\begin{equation}
\mathcal{M}_c = (m_1 m_2) ^{3/5}/(m_1 + m_2)^{1/5}.
\label{eq:Mc}
\end{equation}
Including the contribution of tides and assuming that the
WD spins are synchronized with the orbital period, the rate of change of
\emph{orbital} frequency changes according to \citep{2011ApJ...740L..54B}:
\begin{equation}
\dot{f} = \dot{f}_{0} \: (1\: +5\:\Delta_{Q}  + 3\:\Delta_{I}  ), 
\label{eq:fdot_tides}
\end{equation}
In the equation above, 
\begin{equation}
\Delta_{Q}  = \frac{Q\: (\pi\:f)^{4/3}}{G^{2/3}M^{5/3}}, \:\:\: \Delta_{I}  = \frac{(I_1\:+\:I_2)\: (\pi\:f)^{4/3}}{\mu  G^{2/3}M^{2/3}}
\label{eq:del_Q}
\end{equation}
where $Q = k_2I_i$ is the quadrupole moment, $k_2$ describes
the structure of the star and $I_i = m_1r_1^2$ is the moment of inertia
of each star (with radius $r_i$). This can be translated in terms of
$\mathrm{P_{orb}}$, $\omega$, or $f_{\mathrm{EM}}$ via: $\omega =
2\pi/\mathrm{P_{orb}} = 2\pi f_{\mathrm{EM}} = \pi f$. Thus, including
the orbital decay due to tides the GR formulation in Eq. 3 will then
change according to:
\begin{equation}
\left(\frac{\ddot{f}\:f} {{\dot{f}}^2}\right)_{\mathrm{tides+GW}} =
\frac{\frac{11}{3} \: + \: 25 \Delta Q \: + \: 15 \Delta I } {1 \: + \: 5 \Delta Q \: + \: 3 \Delta I }, 
\label{eq:tot_fdot}
\end{equation}
Given the measured masses, radii and the present orbital period (or
equivalently GW $f$) of J0651 and the assumptions from
\citep{2011ApJ...740L..54B}, we get $\left((\ddot{f}\:f)/
  {\dot{f}}^2\right)_{\mathrm{tides+GW}}  = 3.73138$. This is a deviation from GR driven case of $11/3$ by 
only $1.7650\%$\footnote{This
estimate depends strongly on the moment of Inertia $I_i$ of each of
the binary masses; in fact the term $I_2$ (i.e. of the lower of the
masses which is more tidally deformed by the more massive mass)
derived from a model for a tidally deformed star is the term that most
affects the ratio in Eq.\,\ref{eq:tot_fdot}}. In
deriving this value we only accounted for the lower mass white dwarf
which is distorted whereas the higher mass white dwarf is relatively
undistorted and thus its quadrupole moment can be 
ignored. The deviation above implies that the
measured quantities from which $y$ is derived should have accuracies 
at the level of less than a few percent in order to distinguish tidal
dissipation from GW radiation in J0651-like systems.

In Figure~\ref{fig:GW_errors} fractional accuracies $f, \dot{f},
\ddot{f}, y$ are plotted as a function of orbital period for the three
binaries with GW parameter values listed in Table 1. In the
figure, the size of the open and filled circles and the square
represent the S/N of the system at that orbital period (or
equivalently the GW frequency) from the GW observations. These GW
uncertainties decrease with increasing GW frequency as expected since 
they have higher S/N and at high-$f$ the resolution of
the GW parameters decrease as doppler modulation gains significance 
(see discussion in \cite{2013A&A...553A..82S,
  2003CQGra..20S.163C}). The vertical lines in the 
top-left panel from right to left are the lowest limit
of the orbital periods of the high-$f$ system, high-mass system 
and J0651 respectively where the mass transfer will ensue. This is
derived simply by setting the Roche-lobe of donor WD
\citep{1983ApJ...268..368E} equal to the size of its predicted
zero-temperature radius from the mass \citep{1988ApJ...332..193V}. A
more accurate estimate of the period at which mass transfer starts is
obtained by fitting the spectra with the best matching He WD models
and this gives a larger value for the $\mathrm{{P}_{orb}}$, for e.g. for
J0651 the mass transfer will start when it evolves to a period of
$\sim 420$s the \citep{2007MNRAS.382..779P} and making it difficult
to disentangle the tidal effects. In the figure, the accuracies in the
parameters from observing the EM timing measurements for J0651 are
shown for an observation length of 10 years (in diamond) and for 20
years (in plus). The accuracy in $y$ for both the cases of GW and EM
uncertainties is computed using propagation of errors using
Eq.\,\ref{eq:tides}. The timing accuracy is assumed constant for 
all periods and this implies the uncertainties in the phase increase
for smaller periods however, the values of $\dot{f}, \ddot{f}$
increase more steeply and thus we predict \textbf{increasing} accuracies of
$\dot{f}, \ddot{f}$ for smaller periods. It is clear from these
uncertainties that using only GW data measuring a tidal contribution
is only possible if it is huge for a system like J0651 during their
evolution until mass transfer starts. However the EM and GW fractional
uncertainties in $\dot{f}, \ddot{f}$ for the high-$f$ binary are both
very precise at $\sim 10^{-5}, 10^{-3}$ respectively with which a
small deviation in $y$ can be measured. However, the chances of
observing an eclipsing high-$f$ binary and within 1kpc is almost 0 and
thus measuring tides for such a system only with EM is most likely not
possible. 
\begin{figure*}
\centering 
\includegraphics[width=\textwidth]{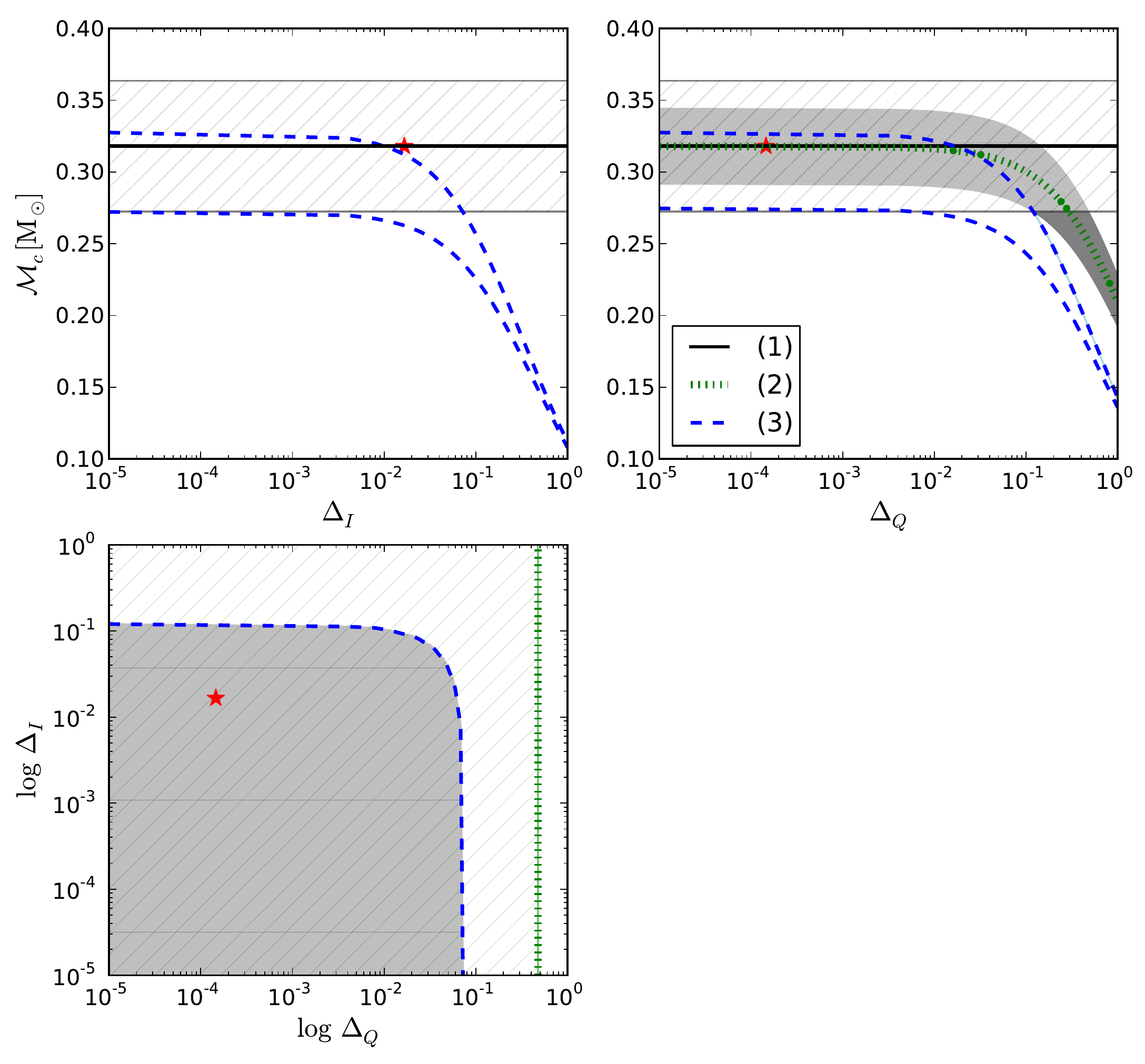}
\caption{Measurability of tidal effects by determining the
\emph{inconsistency} between chirp mass $\mathcal{M}_c$ measurement 
from different methods as a function of the the tidal deviation terms
$\Delta_{I}$ and $\Delta_{Q}$ (characterizing the strength of the
tides) for J0651 system. The methods are: (1) using $m_1$ and $m_2$ in
Eq~\ref{eq:Mc} (2) using $\mathcal{A}$ and $d$ in
Eqs~\ref{eq:amp_del},~\ref{eq:amp}  (3) using $\dot{f}$ in
Eq~\ref{eq:fdot_tides}. Top-left and right panels show  the
$\mathcal{M}_c$ computed from method 1 (shown in (black) 
solid line), from method 2 (in (green) dotted line) and from method 3
(in (blue) dashed lines). Method 1 is not influenced by tides (i.e. no dependence
on $\Delta_{I}, \Delta_{Q}$) whereas method 2 depends on $\Delta_{Q}$
only and method 3 depends on both $\Delta_{I}$ and $\Delta_{Q}$. The
$1-\sigma$ uncertainties in the $\mathcal{M}_c$ for method 1 are from
$\sigma_{m_1}, \sigma_{m_2}$ shown in (grey) hatched area; for method
2 are from $\sigma_{\mathcal{A}}, \sigma_{d}$ shown in grey filled 
area (in top-right panel); for method 3 is from $\sigma_{\dot{f}}$ is
not visible since the relative uncertainty is $\sim 10^{-3}$. In the
top-left panel, the top and bottom (blue) dashed curves correspond to two 
values of $\Delta_{Q} =10^{-5}, 0.0683$ and in the top-right panel the same correspond
to two values of $\Delta_{I}=10^{-5}, 0.1205$ as $\mathcal{M}_c$ measurement via
$\dot{f}$ (method 3) depends on both the tidal deviation
terms. Observe that the three methods (in top panels) show 
inconsistency in chirp mass  with increasing $\Delta_{I}, \Delta_{Q}$,
however the measurement uncertainties are too large in order to
measure the inconsistency for small $\Delta_{I}, \Delta_{Q}$.  The deviations at
which the inconsistency can be measured within the uncertainties are
determined by where the (blue) dashed lines and (grey) shaded area 
cross with the (grey) hatched area in the top panes.  In the bottom
panel these crossings are shown as a joint boundary in $\Delta_{I}$
and $\Delta_{Q}$. The estimated $\Delta_{I}, \Delta_{Q}$ for J0651 
system are marked by the red star. It shows that
constraining the tidal deviation terms is not feasible (for J0651-like
systems) because typically the values of
$\Delta_{I}$ and $\Delta_{Q}$ are smaller than the measurement
uncertainties in $m_1$, $m_2$, $\mathcal{A}$, and $d$.}
\label{fig:chirp_mass}
\end{figure*}
\section{Combining EM $\dot{f}$ and GW measurements}
We find that from the timing measurements with 20
year duration the orbital decay will be observed with fractional
accuracies with up to 5 orders of magnitude better than the GW
accuracy for a system like J0651. Coincidentally a timing length of 20
years coincides with eLISA's launch giving us an opportune time to
combine the EM measurements with the GW ones in improving our
knowledge of J0651-like system parameters. In this section we address
to what extent we can measure the tidal deviation terms introduced in
Sect. 4. We also address how the knowledge of $\dot{f}$ improves the
measurement of J0651's physical parameters of astrophysical interest
such as the masses, inclination and the distance to the source. 
\subsection{Constraining the tidal deviation terms, $\Delta_Q$, $\Delta_I$} 
The measurement of $\dot{f}$ can put constraints on the
tidal contributions. Here we explore these constraints 
formulated in \cite{2011ApJ...740L..54B} that are expressed as
$\Delta_Q$ and $\Delta_I$ in Eq~\ref{eq:fdot_tides}. Under the same
formulation, the GW amplitude that takes into account the quadrupole
correction to the potential of the tidally distorted primary mass
(less massive of the two) can be expressed as:
\begin{equation}
\mathcal{A} = \mathcal{A}_\mathrm{o} (1+\Delta{Q}), 
\label{eq:amp_del}
\end{equation} 
where the GR driven GW amplitude is given by: 
\begin{equation}
\mathcal{A}_\mathrm{o} =\frac{4(G\mathcal{M}_c)^{5/3}}{c^4 d}\left(\pi f \right)^{2/3}
\label{eq:amp}
\end{equation} 

Assuming binary evolution is only driven by gravitational waves, we
compute chirp mass in three ways for J0651 system: (1) measurements of
$m_1$ and $m_2$ ($\mathcal{M}_c$, Eq~\ref{eq:Mc}), (2) measurements of
$\mathcal{A}$ and $d$ ($\mathcal{M}_c(\mathcal{A}_{\mathrm{o}}, d)$,
Eqs~\ref{eq:amp_del},~\ref{eq:amp}) and (3) measurement of $\dot{f}$
($\mathcal{M}_c(\dot{f}_{\mathrm{o}})$, Eq~\ref{eq:fdot_tides}).  The 
uncertainties in the measurements of $m_1$, $m_2$ and $d$ are taken to
be $10\%$ for the masses and distance. Uncertainty in amplitude is
taken from the FIM matrix $\sim 10\%$ for eclipsing J0651 and the
uncertainty in $\dot{f}$ is taken to be $0.01\%$, a conservative
estimate from Figure~\ref{fig:GW_errors}. For the three estimates of
chirp masses we compare for what values of $\Delta_Q$ and $\Delta_I$
are the  $\mathcal{M}_c(\mathcal{A}_{\mathrm{o}}, d)$ and
$\mathcal{M}_c(\dot{f}_{\mathrm{o}})$ inconsistent with
$\mathcal{M}_c$. In the top panels of
Figure~\ref{fig:chirp_mass} $\mathcal{M}_c$ is shown in black line
with the corresponding $1-\sigma$ uncertainty shown by (grey) hatched
area. This estimate of $\mathcal{M}_c$ does not depend on $\Delta_Q,
\Delta_I$. Since $\mathcal{M}_c(\dot{f})$ depends on both $\Delta_Q,
\Delta_I$, it is plotted for two values of $\Delta Q = [10^{-5},
0.12]$ shown in upper and lower (blue) dashed lines respectively in
the top-left panel. In the top-right panel $\mathcal{M}_c(\dot{f})$ is
shown for two values of $\Delta I = [10^{-5}, 0.068]$ corresponding to
the upper and lower (blue) dashed lines respectively. The
predicted deviations from average measurements of the masses and radii
for J0651 are marked by the (red) star. \cite{2011ApJ...740L..54B} 
estimate that $\Delta_Q, \Delta_I = 1.46\times10^{-4}, 0.0166$ for
J0651. The relative uncertainties of $\mathcal{M}_c(\dot{f})$ are in
the level of $10^{-3}$ not visible in the figure. Finally 
$\mathcal{M}_c(\mathcal{A}, d)$ can constrain $\Delta Q$ only and it 
is shown in dotted line with uncertainties in grey shaded area in the
middle panel. The range of values of $\Delta_Q$ and $\Delta_I$ for
which the three sets of chirp masses are inconsistent with each other
within their uncertainties can be read from the figure which are,
$\Delta_I > 0.120, \Delta_Q > -0.478$. In the bottom panel
the constraints in $\Delta_Q$ and $\Delta_I$ using both EM and GW data
are shown by the (blue) dashed curve (via method 2) and (dotted)
vertical line (via method 3). From the bottom panel it can be
seen that measuring tidal deviation $\Delta_Q$ is not feasible within
the uncertainties in $\mathcal{A}, d$ marked by the (green) hatched
area. Also measuring the deviation term $\Delta_I$ which is larger (at
$\lesssim 10^{-2}$) is not feasible within the uncertainties in $m_1,
m_2$ marked by the (grey) shaded area. Even though we expect strong
tidal influence in detached white-dwarf systems such as J0651,
measuring that contribution is unlikely unless the 
the measurements in GW amplitude, distance or the individual masses
should be also in the order of $\lesssim 10^{-2}$ for J0651-like systems.
We conclude that tidal physics can be studied for 
high-mass binaries at opportune frequencies which implies larger
values of decay rate measurable from the GW data. 
\subsection{Constraining the binary parameters} 
In our earlier work (Shah \& Nelemans \emph{subm.}, SN2013, hereafter) we
studied the effect of combining 
GW and EM observations, where we considered the following EM
measurements: the $d$ from \emph{Gaia} satellite, primary mass
$m_1$ from spectroscopy, radial velocity $K_1$ also from spectroscopy
and possibly inclination $\iota$ from the fact that the binary can be
eclipsing. We found that adding one or more of these measurements
significantly improves our knowledge of the unknown astrophysical
parameters of the binary and the improvement depends on the
inclination of the source. In this study we add the EM information of
the orbital decay rate $\dot{P}_{\mathrm{orb}}/\dot{f}$ (from Sect. 4)
to the above list of EM observations and study if and how much this
improves the binary parameters secondary mass $m_2$ and distance
$d$ compared to scenarios considered in SN2013. The uncertainties in
$m_1, K_1, d$ are taken to be $10\%$ as explained in SN2013,
whereas $\dot{f}$ is taken to have an accuracy of
$\sim 0.1\%$ as measured from the timing eclipses J0651 (see
Sect. 4). Our method of combining each set of EM data with that of
the GW data (i.e. amplitude $\mathcal{A}$ and inclination) is described in SN2013
and here we will summarize the advantage of including
$\dot{P}_{\mathrm{orb}}$ for each of the scenarios discussed in the
earlier paper. Each of the scenarios below 
include GW measurements $\mathcal{A}$, $\iota$ of J0651 system as a
function of its inclination. We also assume the GW frequency of the
source is known exactly since its relative uncertainty from GW
observation for J0651 system is $10^{-7}$Hz.
\begin{figure*}
\centering 
\includegraphics[width=\textwidth]{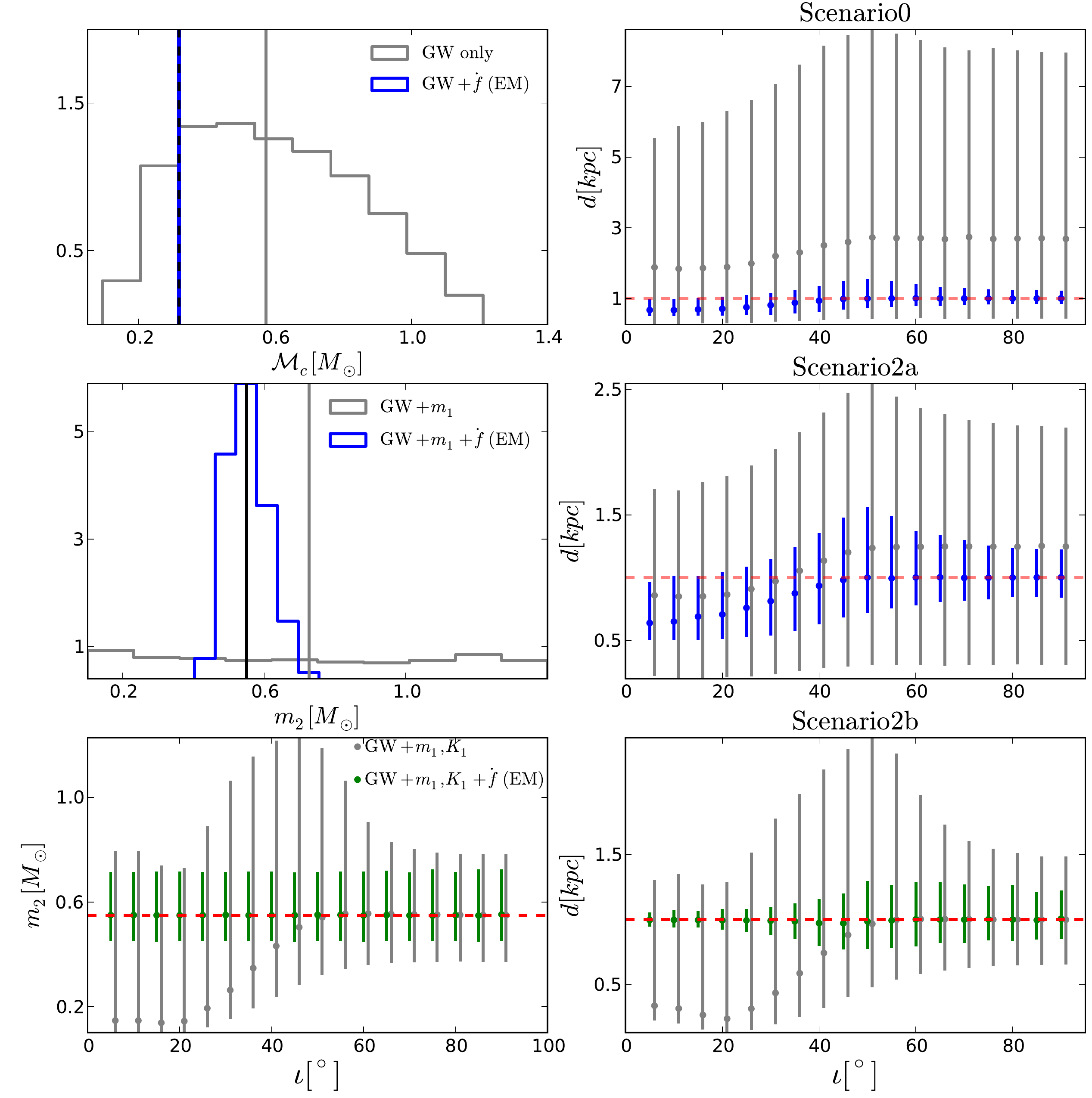}
\caption{Constraints in the binary parameters of J0651 given
by complementing GW observations with EM data for three scenarios. Top 
row: \emph{Scenario 0}, comparison of $\mathcal{M}_c$ and 95
percentile   uncertainties in $d$ as a function of inclination with EM
data on $\dot{f}$ (shown in blue) versus GW data only (shown in
grey). In the top-left and top-middle panels, the vertical (grey and
blue) lines are distribution medians and vertical dashed line is the
\emph{real} value of the system. In all the right panels and
bottom-left panel, the (red) horizontal is the \emph{real} value of
the source parameter. Middle row: \emph{Scenario 2a}, comparison of
$m_2$ and $d$ with EM data on $\dot{f}, m_1$ (shown in blue) versus GW
data + $m_1$ only (shown in grey). Bottom row: \emph{Scenario 2b},
comparison of the same with added information of $K_1$ for both cases
which are shown in green. As expected adding $\dot{f}$ improves the
distance estimates significantly in all three cases when compared to
the corresponding cases address in SN2013.}
\label{fig:case3}
\end{figure*}
\begin{enumerate}
\item{Scenario 0: GW data + $\dot{f}$ vs.
  GW data only}
In SN2013 we found that distance can be estimated using GW
amplitude. The chirp mass in this case was simply estimated for WDs
using uniform distributions of the masses ($m_i \: \epsilon  \:[0.1, 1.4]
M_{\odot}$) which is shown in grey in the top-left panel of
Figure~\ref{fig:case3}. The 95 percentile in distances as a
function of inclination are shown in the bottom-left panel in
grey. Adding EM data of $\dot{f}$ with $0.1\%$ accuracy will constrain
the 95 percentile in $\mathcal{M}_c$ to a much better accuracy of
$0.11\%$ compared to the SN2013 which is shown in blue in the top-left
panel of Figure~\ref{fig:case3}. The medians of these distributions
are shown 
in solid lines and the real value is shown in dashed black line. Hence
the distances can be also constrained to much better accuracies shown
in blue in the bottom-left panel where we find that the (relative) 95
percentile uncertainties in distances range from $36\%$ to $19\%$ for
inclinations of $5^{\circ}$ (face-on orientation) to $90^{\circ}$
(edge-on orientation) which are significantly better compared to the
grey lined found in SN2013. 
\item{Scenario 1: GW data + $\dot{f}, d$ vs. GW data + $d$}
\newline
In SN2013 we found that using distance $d$ and $\mathcal{A}$ we could
estimate $\mathcal{M}_c$ as a function of inclination where the 95
percentile in $\mathcal{M}_c$ fared better at higher inclinations with
$\sim 16\%$ and worse at lower inclinations. As shown above, adding EM data
from $\dot{f}$ already constrains $\mathcal{M}_c$ much better to
$0.11\%$ for all inclinations. Thus adding EM data on $d$ does not add
much unless both $d$ and $\mathcal{A}$ are known to better accuracies
that $\dot{f}$.
\item{Scenario 2a: GW data + $\dot{f}, m_1$ vs. GW data + $m_1$}
In SN2013 we found that combining EM data on $m_1$ with GW
$\mathcal{A}$ provided an estimate of the secondary mass $m_2$ and
constraints on the distance as a function of inclination. The
distribution of $m_2$ (which is simply solved using grey distribution
in $\mathcal{M}_c$ in the top-left panel) is shown also in grey in the
top-middle panel of the figure. The 95 percentiles in $d$ using this
$m_2$ and $\mathcal{A}$ are shown in grey in the bottom-middle
panel. Adding EM data of $\dot{f}$ will improve the accuracy of $m_2$
reducing the 95 percentile uncertainty to $25\%$
accuracy owing to a very accurate $\mathcal{M}_c$ (as
discussed above). This reduced distribution in this $m_2$ is shown in blue in the
top-middle panel. This $m_2$ in combination with 
the GW $\mathcal{A}$ constrains the distance with better accuracies
compared to SN2013 whose 95 percentiles are also shown as a function
of inclination in blue lines in the bottom-left panel of the
Figure. Thus adding $\dot{f}$ in this scenario improves $m_2$ and $d$
significantly.
\item{Scenario 2b: GW data + $\dot{f}, m_1, K_1$ vs. GW data + $m_1, K_1$} 
In SN2013 we found that combining single-line spectroscopic data,
i.e. $m_1, K_1$ with the GW $\mathcal{A}$ constrained both the $m_2$
and $d$ as a function of inclination whose 95 percentiles are shown in grey in
top-right and bottom-right panels of Figure~\ref{fig:case3}
respectively.  Here we find that adding $\dot{f}$ will improve both of
these quantities significantly whose respective 95 percentiles are shown in green
lines. To explain these improvements we briefly explain how these
quantities are estimated. As explained in the case above we have an accurate
constraint on $m_2$ using $m_1$ and $\mathcal{M}_c$. Using the GW
inclination and the masses we compute the radial velocity at each
inclination, $K_{\mathrm{GW}}$. At each inclination $K_{\mathrm{GW}}$
is compared against the measured $K_1$. Using the observed
distribution selecting a subset of $K_{\mathrm{GW}}$ with a
probability distribution of $K_1$ constrains a subset in the rest of
the parameters: $m_1,m_2, \mathcal{A}, d$ even further. The
reduced uncertainties in $m_2, d$ calculated in this way are are shown in green
in the top-right and bottom-right panels of the Figure. The method
described is akin to Scenario 2c discussed in SN2013 in detail. Thus
adding EM data of $\dot{f}$ to $m_1, K_1$ improves the distance
estimates significantly especially at lower inclinations. We find the
(relative) 95 percentile in $m_2$ range from $25\% - 17\%$ for face-on
to edge-on systems. And the same for $d$ range from $6\% - 19\%$. 
\item{Scenario 3: GW data + $\dot{f}, m_{1,2}, K_{1,2}$ vs. GW data + $m_{1,2}, K_{1,2}$} 
In SN2013 we found that combining $m_{1,2}$, $K_{1,2}$ with GW data
improves the $\mathcal{A}, \iota$ especially for lower inclination
systems and this in turn constrains the distance of the binary (to
roughly $30\%$). Those distances can be compared with the independent
estimate of the same using $\dot{f}$ explained above in Scenario
0. Since $m_{1,2}$ are considered to have $10\%$ accuracies much
larger than $0.1\%$ accuracy in $\dot{f}$, the chirp mass is still
better determined in the case where $\dot{f}$ is known and thus adding
information from $m_{1,2}$, $K_{1,2}$ does not improve the constraint
in distance any further.
\end{enumerate}
\section{Conclusion}
We investigated the feasibility of detecting tides in detached
(white-dwarf) binaries from eLISA detector by calculating
uncertainties of the parameters, $\dot{f}, \ddot{f}$ as a function of
the orbital frequency. We implement Fisher-matrix methods to compute
the GW parameters uncertainties and compares them with the accuracies
from the mid-eclipsing timing measurements where the observation
length is taken to be $\ge$10 years. We also study the quantitative
improvements in binary parameters when an EM data on
$\dot{P}_{\mathrm{orb}}$ is combined with GW data and other possible
sets of EM data. From our analyses of J0651 and 
higher mass systems (see Table 1), we conclude: 
\begin{enumerate}
\item Unless eLISA can discover systems like the high-$f$ binary, GW
  data alone will not suffice in measuring $\dot{f}, \ddot{f}$ 
  precisely enough for a system like J0651. However finding
  such high-$f$ binaries near by ($\le 1$kpc) is very unlikely. 
\item Eclipse timing measurements for 10 years for J0651-like
    systems will provide a very precise measurement of $\dot{f}$ to
    less than $1\%$. However, measuring a 2-5$\%$  
    contribution from tides in $\dot{f}$ for such binaries is
    \emph{only} possible if the $\mathcal{M}_c$ and/or $d$ are also
    known to $\sim  1\%$ accuracies. Additionally detecting
    a collective phase shift in the GW phase using \emph{only} GW data for
    J0651 as has been suggested 
  \citep{2012MNRAS.421..426F} is not possible.
\item For systems driven by \emph{only} GW radiation, an EM measurement of
  $\dot{f}$ combined with GW measurement of
  $\mathcal{A}$ provides us a very precise
  measurement of $\mathcal{M}_c$. We compare this to our previous study
  in SN2013 where we computed improvement in binary parameters for the
  case of J0651 as a function of its inclination. We find that
  $\dot{f}$ can constrain $m_2$ and 
  $d$ more accurately when considering various scenarios where EM data on
  $m_1, K_1$ are known. We find that knowing only $\dot{f}$ constrains
  the 1-sigma in $\mathcal{M}_c$ to $0.3179\pm 0.0002 M_{\odot}$. This
  further constrains $d$ from $0.70_{0.46}^{0.95}$kpc (face-on)
  to $1.00_{0.70}^{1.32}$kpc (edge-on). Adding EM data on $m_1$
  constrains the $m_2$ to $0.55_{0.49}^{0.62}M_{\odot}$. Finally
  adding EM data on $m_1, K_1$ constrains distance from
  $1.00_{0.96}^{1.03}$kpc (face-on) to $1.00_{0.91}^{1.10}$kpc 
  (edge-on). We conclude that compared to the scenarios in
  SN2013 our knowledge of the chirp mass, secondary mass and the
  distance improve significantly when the eclipse timing
  measurements in $\dot{f}$ will be included in the GW-EM synergy.
\end{enumerate}
\begin{acknowledgements}
This work was supported by funding from FOM.
\end{acknowledgements}
\bibliographystyle{apj} 
\bibliography{/Users/swetashah/Documents/writings/literature_binary_science,/Users/swetashah/Documents/writings/literature_data_analysis}
\appendix
\section{Variance-covariance matrixes of J0651, and B2}
We have listed the VCM matrices for the binary systems that we used in
our analysis. There are 9 parameters that described them which are
listen in the first row of the matrices below and for each binary, the
values are listed in the row with $\theta_{i}$. The diagonal elements
are the absolute uncertainties in each the 9 parameters and the
off-diagonal elements are the normalized correlations, i.e. $
\mathrm{c}_{ii} = \sqrt{\mathcal{C}_{ii}} \equiv \sigma_{i},$
$\mathrm{c}_{ij}
=\frac{\mathcal{C}_{ij}}{{\sqrt{\mathcal{C}_{ii}\mathcal{C}_{jj}}}}.$
The strong correlations between parameters (i.e. whose
  magnitudes are ≥ 0.7) are marked in bold in the VCMs below.\\ 

VCM 1: J0651, S/N $\sim 13$.  \\
\resizebox{\linewidth}{!}
{
\bordermatrix{~ & \mathcal{A} & \phi_0 & \cos\iota & f & \dot{f}& \ddot{f}&\psi & \sin\beta & \lambda \cr
\theta_i&1.67\times10^{-22} & \pi & 0.007& 2.61\times10^{-3}& -3.35\times10^{-17}&1.57\times10^{-30}&\pi/2&0.01&1.77\cr \hline \cr
\mathcal{A}  &1.586\times10^{-23}& -0.0& \;\;\;0.0& \;\;\;0.01& -0.01& -0.0& \;\;\;0.02& \;\;\;0.03& -0.06\cr
\phi_0& & \;\;\;0.364&-0.01& \mathbf{-0.91}& \;\;\;\mathbf{ 0.82}& -0.01& \;\;\;0.01& \;\;\;0.11& \;\;\;0.08\cr
\cos\iota && &\;\;\;0.044&\;\;\;0.01& -0.01& \;\;\;0.0& -0.01& \;\;\;0.07& -0.33\cr
f & &&& 3.807\times10^{-9}& \mathbf{-0.98}& \;\;\;0.01& -0.01& -0.08& -0.15\cr
\dot{f}& &&&& 1.059\times10^{-16}& -0.04& \;\;\;0.01& \;\;\;0.04& \;\;\;0.19\cr
\ddot{f}& &&&&& 1.047\times10^{-26}& \;\;\;0.0& \;\;\;0.0& \;\;\;0.08\cr
\psi & &&&&&&\;\;\;0.041& -0.02& \;\;\;0.05\cr
\sin\beta &&&&&&& & \;\;\;0.069& \;\;\;0.08\cr
\lambda&&&&&&&&&  \;\;\;0.019 \cr  \cr}
}\\
\\
\\


VCM 3:  high-frequency binary, S/N $\sim135$.\\
\resizebox{\linewidth}{!}
{
\bordermatrix{~ & \mathcal{A} & \phi_0 & \cos\iota & f & \dot{f}& \ddot{f}&\psi & \sin\beta & \lambda \cr
\theta_i&3.698\times10^{-22} & \;\;\;3.666 & -0.331& 17.695\times10^{-3}& 1.988\times10^{-13}&8.191\times10^{-24}&1.97&0.685&5.411\cr \hline \cr
\mathcal{A}  &5.02\times10^{-24}& -0.15& \;\;\;\mathbf{ 0.79}& \;\;\;0.05& -0.05& \;\;\;0.05& \;\;\;0.28& \;\;\;0.29& -0.21\cr
\phi_0& (0.0136)& \;\;\;0.048&-0.07& \;\;\;\mathbf{0.87}& \;\;\;\mathbf{0.82}& \mathbf{-0.76}& -0.36& -0.26& 0.02\cr
\cos\iota && &\;\;\;0.008&\;\;\;0.07& -0.06& \;\;\;0.05& \;\;\;0.04& \;\;\;0.39& -0.07\cr
f & &&& 8.228\times10^{-10}& \mathbf{-0.98}& \;\;\;\mathbf{0.92}& -0.02& \;\;\;0.24& \;\;\;0.22\cr
\dot{f}& &&&(4.65\times10^{-8})& 5.169\times10^{-17}& \mathbf{-0.98}& -0.0& -0.27& -0.17\cr
\ddot{f}& &&&&(2.6\times10^{-4})& 1.4476\times10^{-24}& \;\;\;0.02& \;\;\;0.30& \;\;\;0.14\cr
\psi & &&&&&(0.176)&\;\;\;9.86\times10^{-3}& -0.09& -0.58\cr
\sin\beta &&&&&&& & \;\;\;2.5\times10^{-4}& \;\;\;0.14\cr
\lambda&&&&&&&&&  \;\;\;4.1\times10^{-4}\cr  \cr}
}
\end{document}